%% file: main.tex
\documentclass[twocolumn]{aastex62} % preprint2

\include{definitions}

\received{May 4, 2019}
\revised{June 4, 2019}
\accepted{\today}
\submitjournal{ApJL}

\shorttitle{Spin of M87*}
\shortauthors{Nemmen}

\begin{document}

\title{The Spin of M87*}

\email{rodrigo.nemmen@iag.usp.br}

\author{Rodrigo Nemmen}
\affil{Universidade de S\~ao Paulo, Instituto de Astronomia, Geof\'{\i}sica e Ci\^encias Atmosf\'ericas, Departamento de Astronomia, S\~ao Paulo, SP 05508-090, Brazil}
\nocollaboration

\begin{abstract}
Now that the mass of the central black hole in the galaxy M87 has been measured with great precision using different methods, the remaining parameter of the Kerr metric that needs to be estimated is the spin $a_*$. We have modeled measurements of the average power of the relativistic jet and an upper limit to the mass accretion rate onto the black hole with general relativistic magnetohydrodynamic models of jet formation. This allows us to derive constraints on $a_*$ and the black hole magnetic flux $\phi$. We find a lower limit on M87*'s spin and magnetic flux of $|a_*| \geq 0.4$ and $\phi \gtrsim 6$ in the prograde case, and $|a_*| \geq 0.5$ and $\phi \gtrsim 10$ in the retrograde case, otherwise the black hole is not able to provide enough energy to power the observed jet. These results indicate that M87* has a moderate spin at minimum and disfavor a variety of models typified by low values of $\phi$ known as ``SANE'', indicating that M87* prefers the magnetically arrested disk state. We discuss how different estimates of the jet power and accretion rate can impact $a_*$ and $\phi$.
\end{abstract}

\keywords{accretion, accretion disks -- black hole physics -- galaxies: active -- galaxies: individual (M87) -- galaxies: jets }

\section{Introduction} \label{sec:intro}

The shadow cast by the event horizon of a black hole (BH) has been imaged for the first time with the Event Horizon Telescope (EHT) for the supermassive black hole (SMBH) at the center of the galaxy M87, known as M87* (\citealt{EHTC2019}, hereafter EHTC1). The very long baseline interferometry (VLBI) observation at a wavelength of 1.3 mm of the asymmetric bright emission ring gives an angular diameter of $d=42 \pm 3 \ \mu$as, which allows an unprecedented constraint on the SMBH mass of $M=(6.5 \pm 0.7) \times 10^9 M_\odot$--the first fundamental parameter of the \cite{Kerr1963} metric. The inferred mass is in agreement with--and hence strongly favors--the mass measurement based on stellar dynamics \citep{Gebhardt2011}. 

However, the second parameter of the Kerr metric--the dimensionless spin $a_* \equiv Jc/GM^2$ where $J$ is the angular momentum of the black hole--is much harder to constrain using only shadow observations. One of the reasons is that the ring diameter has a very weak dependence on $a_*$ and the disk inclination, varying by only 4\% in the range $a_*=0$ to $\approx 1$ \citep{Takahashi2004,Johannsen2010}. 

Besides mass and spin, black hole accretion is also described by two other parameters: the mass accretion rate onto the BH $\dot{M}$ and the magnetic flux $\Phi$ crossing one hemisphere of the event horizon. 
Based on the properties of the asymmetric ring observed in 2017 with the EHT as well as other constraints, \cite{EHTC2019d} (hereafter EHTC5) obtained some bounds on $\Phi$ and $a_*$. Models with high, retrograde spins and high $\Phi$ are rejected, as well as $a_*=0$ models which fail to provide enough power to the jet. 

In this Letter, we set further bounds on the values of $a_*$ and $\Phi$ of M87* by modeling the energetics of its relativistic jet as being powered by the Blandford-Znajek process--the extraction of BH spin energy through electromagnetic torques. The observational input to our estimates are the measurements of $M$, $\dot{M}$ and the power carried by the jet. We assume a distance to M87* of 16.8 Mpc \eg{Blakeslee2009, EHTC2019e}. At this distance, cosmological effects are negligible.

\section{Observations} \label{sec:obs}

In this work, the fundamental quantity needed in order to constrain the BH spin in M87* is the efficiency of jet production $\eta \equiv P/ \dot{M} c^2$, where $P$ is the jet power and $\dot{M}$ is the mass accretion rate onto the SMBH. Therefore, $P$ and $\dot{M}$ are the M87* observables that we need in order to apply our models of jet production to constrain the spin parameter.

There are different ways of measuring the jet power of M87, with  different methods giving powers in the range $P \sim 10^{42}-10^{45} \ {\rm erg \ s}^{-1}$ (e.g., \citealt{Reynolds1996, Allen2006, Abdo2009, Degasperin2012, Nemmen2014}; EHTC5 and references therein). Here, we use the jet-inflated X-ray cavities observed in the central regions of M87 with the \textit{Chandra} X-ray Telescope as calorimeters to estimate the jet power \citep{Russell2013}. The jet power was estimated as $P = E_{\rm cav}/t_{\rm age}$, where $E_{\rm cav}$ is the energy required to create the observed cavities and $t_{\rm age}$ is the age of the cavity. The usual assumption in deriving $E_{\rm cav}$ is that the cavities are inflated slowly such that $E_{\rm cav}=4PV$ where $P$ is the thermal pressure of the surrounding X-ray emitting gas, $V$ is the volume of the cavity and the cavity is assumed to be filled up with relativistic plasma. %The age of the cavity is usually assumed to be either the sound-crossing timescale $t_{\rm c_s}$ where $D$ is the distance of the bubble centre from the black hole and $c_s$ is the adiabatic sound speed, or the buoyancy timescale $t_{\rm buoy}=D/v_t$ where $v_t$ is the bubble terminal velocity. 
This method of measuring $P$ is well-established and robust \eg{Birzan2004, Dunn2004, McNamara2012, Hlavacek2013}, giving the jet power averaged over the timescale during which the central engine produces one continuous pair of jets ($\sim 10^6$ years for M87; \citealt{Allen2006}). We believe this is the most direct way of measuring the jet power and therefore accept the X-ray cavity power at face value as M87's jet power, $\log P = 42.9^{+0.27}_{-0.2} \ {\rm erg \ s}^{-1}$. 

For the mass accretion rate onto the BH, we use the constraint obtained by \cite{Kuo2014} based on the Faraday rotation measure (RM) observed with the Submillimeter Array. Kuo et al. measured the M87*'s RM at four  frequencies around 230 GHz to be in the range $−7.5 \times 10^5-3.4 \times 10^5 \ {\rm rad \ m}^{-2}$. By making reasonable assumptions, Kuo et al. estimated the upper limit $\dot{M} (42 r_g) \leq 9.2 \times 10^{-4} \ M_\odot \ {\rm yr}^{-1}$ where $\dot{M} (42 r_g)$ is the mass accretion rate at a distance of 42 gravitational radii from the SMBH ($r_g \equiv GM/c^2$). The corresponding accretion power is $\log (\dot{M} (42 r_g) c^2) = 43.7 \ {\rm erg \ s}^{-1}$. The question is of course how to connect this measurement with the accretion rate $\dot{M}$ near the event horizon. We will come back to this question in section \ref{models}.

\section{Models}    \label{models}

The leading idea to understand how accreting Kerr BHs produce relativistic jets is the Blandford-Znajek process \citep{Blandford1977, Blandford2019}. According to this mechanism, the rotating event horizon is threaded by large-scale magnetic field lines, which were brought in by accreted gas. %The BH exerts a torque on the field lines and progressively loses rotational energy to its environment. This spin energy is converted to a Poynting flux-dominated outflow \eg{McKinney2004,DeVilliers2005} which eventually transfers the electromagnetic energy to particles in the relativistic jet. 
The BH exerts a torque on the field lines and progressively transfers its rotational energy to the relativistic jet \eg{McKinney2004,Semenov2004}. According to the Blandford-Znajek model, the jet power depends on $a_*$ and the magnetic flux $\Phi$ threading the horizon as $P \propto ( a_* \Phi/M )^2$ to first order \citep{Blandford1977}. In reality, for rapidly spinning BHs the jet power depends in a more complicated nonlinear fashion on the spin as higher-order spin corrections become important \eg{Sasha2010}. 

\begin{figure}[h]
\centering
\includegraphics[width=\linewidth]{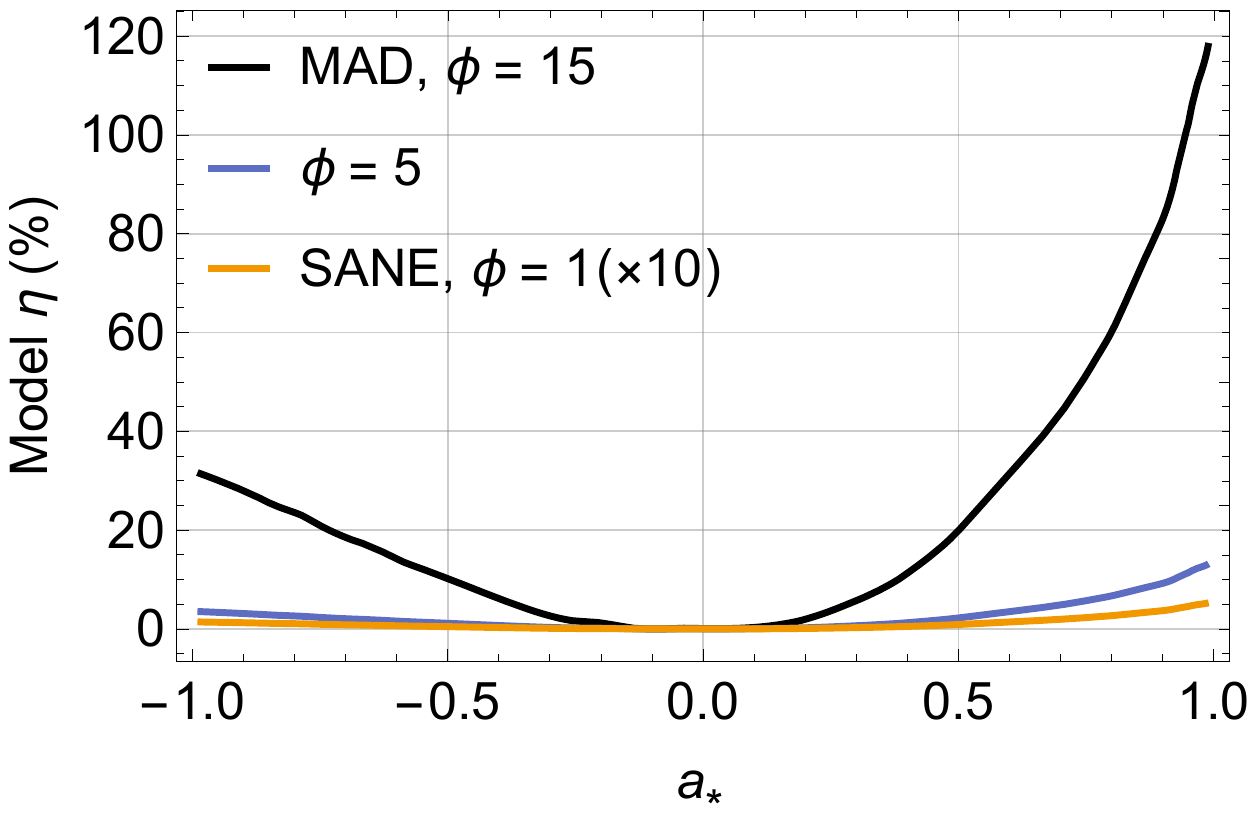}
\caption{The efficiency of the jet production efficiency $\eta$ as a function of the black hole spin $a_*$ from GRMHD simulations of RIAFs, for different values of the magnetic flux $\phi$. Both prograde ($a_*>0$) and retrograde ($a_*<0$) cases are encompassed. The SANE model $\eta$-values were multiplied by 10. }
\label{spin-model}
\end{figure}

We use the results of global, general relativistic magnetohydrodynamic (GRMHD) simulations of radiatively inefficient accretion flows (RIAFs; \citealt{Yuan2014}) around Kerr BHs to model the dependence of the jet power on $a_*$ and $\Phi$. Concretely, we model the jet power as $P = \eta(a_*, \Phi) \dot{M} c^2$ where the jet production efficiency 
\begin{equation}
\eta = \left( \frac{\phi}{15} \right)^2 f(a_*)
\end{equation}
is a function of both the spin and dimensionless magnetic flux $\phi \equiv \Phi/(\dot{M} r_g^2)^{1/2}$. We use the GRMHD results of \cite{Sasha2012a} which are based on the \code{HARM} code \citep{Gammie2003} to set the spin-dependence of $\eta$. \cite{Sasha2012a} carried out RIAF simulations in the ``magnetically arrested disk'' (MAD) limit, for which the magnetic flux saturates at $\phi \sim 15$ \citep{Narayan2003, Sasha2011}, with $h/r \approx 0.3$ where $h$ is the disk thickness. In order to consider the full range of astrophysically relevant magnetic fluxes, we also take into account the case of  ``standard and normal disk evolution'' (SANE) with $\phi \sim 1$ \eg{Narayan2012}. Figure  \ref{spin-model} illustrates the spin-dependence of $\eta$. Notice that this model encompasses both the prograde and retrograde cases in which the disk and BH are rotating in the same and opposite senses, respectively. Retrograde BHs produce less powerful jets \citep{Sasha2012}. 

In order to connect the accretion rate at $r=42 r_g$ constrained by the Faraday RMs of \cite{Kuo2014} with the BH rate, we adopt the simple radial scaling $\dot{M}(r) = \dot{M_0} (r/r_0)^s$ originally proposed as an ansatz by \cite{Blandford1999} in their ``ADIOS'' model--and later supported by many global simulations \eg{Yuan2015}. This $\dot{M}(r)$ scaling corresponds to a density radial profile $\rho(r) \propto r^{-\beta}$ where $\beta = 3/2-s$. We define the BH accretion rate as $\dot{M}(6 r_g)$ so we fix $r=6r_g$, $r_0=42 r_g$ and $\dot{M}_0$ as the accretion rate constrained by Kuo et al. We want to be agnostic regarding the variety of possible density profiles in M87*, therefore we allow $\beta$ ($s$) to vary in the range $1.5-0.5$ ($0-1$), i.e. allowing for different levels of mass-loss in the RIAF. We should note that \cite{Russell2018}  measured $\beta \approx 1.5$ at $r=(0.1-1)$ kpc in M87, which is outside but very close to the Bondi radius ($r_B = 0.03$ kpc).

\section{Results} 

\begin{figure}[h]
\centering
\includegraphics[width=\linewidth]{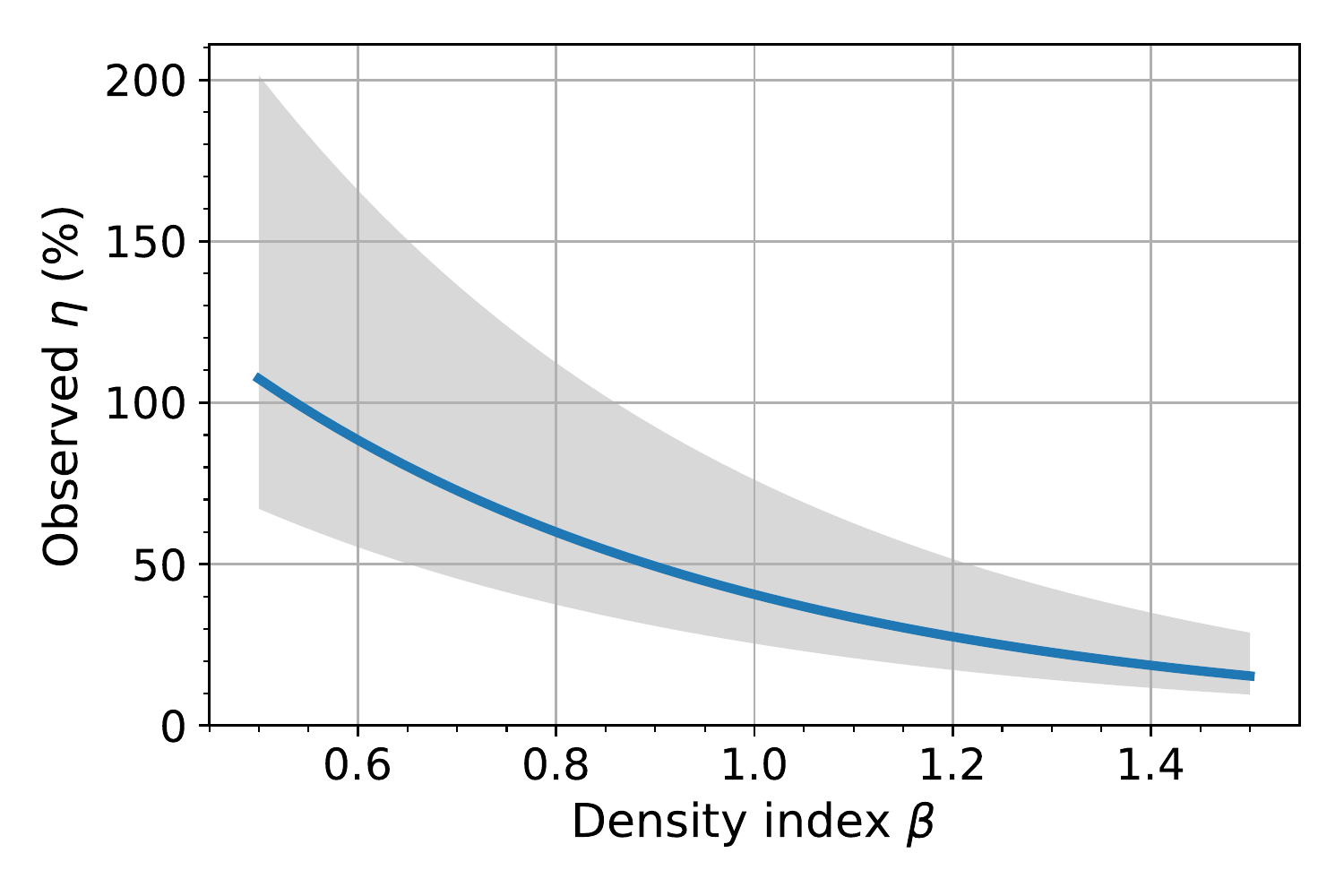}
\caption{Observed jet production efficiency $\eta_{\rm observed}$ for M87* as a function of $\beta$ ($\rho \propto r^{-\beta}$). The shaded area represents the $1\sigma$ uncertainty propagated from the uncertainty in the observed jet power. }
\label{obs-eta}
\end{figure}

Our first result is a model-independent estimate of the jet production efficiency from the SMBH in M87* from the observed jet power and mass accretion rate. Figure \ref{obs-eta} shows this result allowing a variety of density profiles, with $\eta$ varying from $\approx 10\%$ ($1\sigma$ lower limit, $\beta=1.5$) up to about $200\%$ ($1\sigma$ upper limit, $\beta=0.5$) if the density profile flattens towards the BH.

\begin{figure}[h]
\centering
\includegraphics[width=\linewidth]{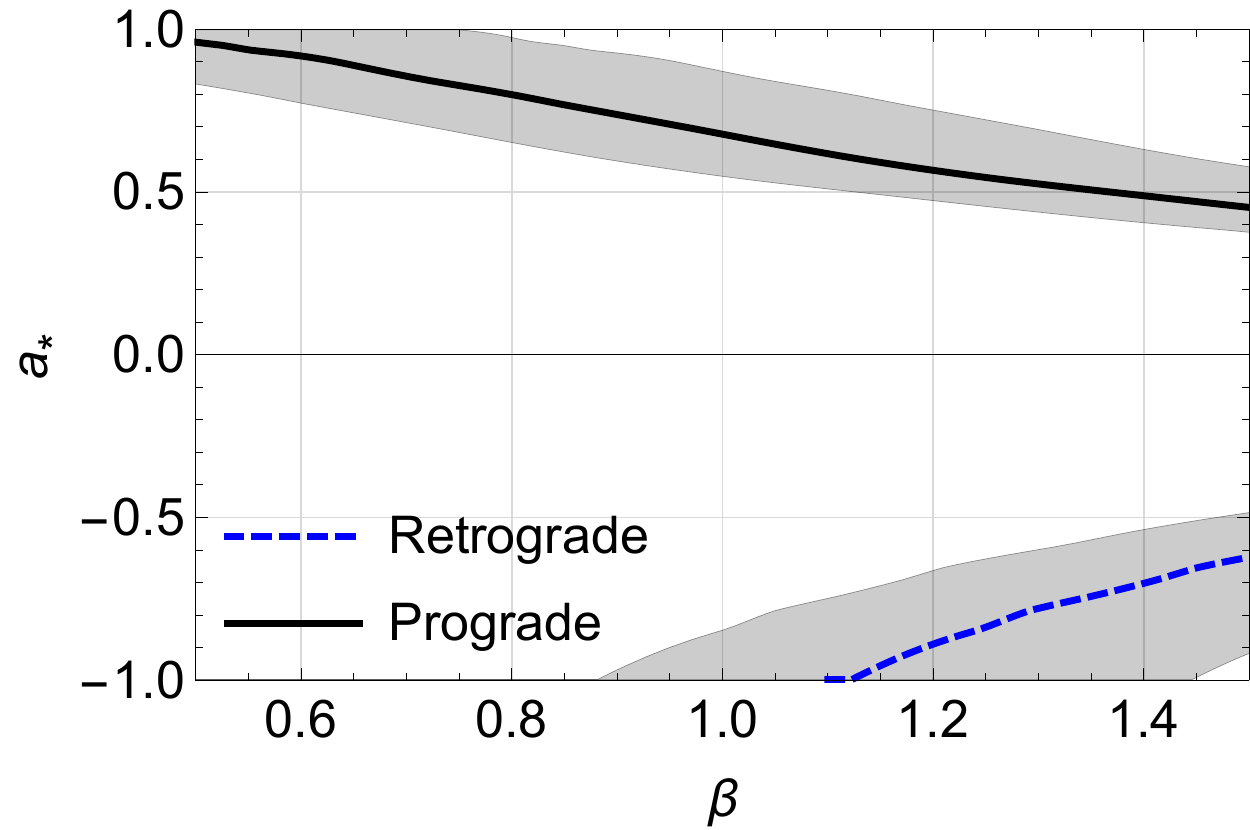}
\caption{The spin of the SMBH in M87* as a function of $\beta$ for both the prograde and retrograde cases, assuming that it is in the MAD state. The shaded region corresponds to the $1\sigma$ uncertainty band around the mean. The uncertainty was propagated from the uncertainty in the observed jet powers. }
\label{spins}
\end{figure}

With the considerations in the previous section, we have a model that provides a full mapping of the observed jet efficiency derived above to the spin and magnetic flux 
\begin{equation}    \label{eq}
\eta_{\rm observed}=\eta(a_*, \phi).
\end{equation}
We proceed by solving this nonlinear equation, using the values of $\eta_{\rm observed}$ displayed in Figure \ref{obs-eta} to constrain the physical parameters of the event horizon beyond its mass--assuming that the Kerr metric is the correct description of the spacetime. Figure \ref{spins} shows the inferred spin of M87* on the assumption that the SMBH is in the MAD state--i.e. with the maximum value of $\phi$. 

The lessons behind Fig. \ref{spins} are the following: (i) If M87* is in the MAD state then the only allowed spins are $a_* \lesssim -0.5$ or $a_* \gtrsim 0.4$ (within the $1\sigma$ uncertainty bands). These are effectively lower limits on $a_*$. (ii) If the density profile of the accretion flow follows $\rho \propto r^{-1.5}$ as in RIAF models without mass-loss \eg{Narayan1994}, then $a_* = 0.45^{+0.12}_{-0.08}$ (prograde) or $a_* = -0.62^{+0.14}_{-0.30}$ (retrograde). (iii) If the BH is retrograde then values of $\beta \geq 0.8$ are favored otherwise M87* would not be able to power its jet through the Blandford-Znajek process; all values of $\beta$ are allowed if the BH is prograde.
Notice that we limit the upper value of $|a_*|$ to one because the BH solution is not valid anymore above this limit.

\begin{figure*}
    \centering
    \begin{minipage}{0.45\textwidth}
        \centering
        \includegraphics[width=\textwidth]{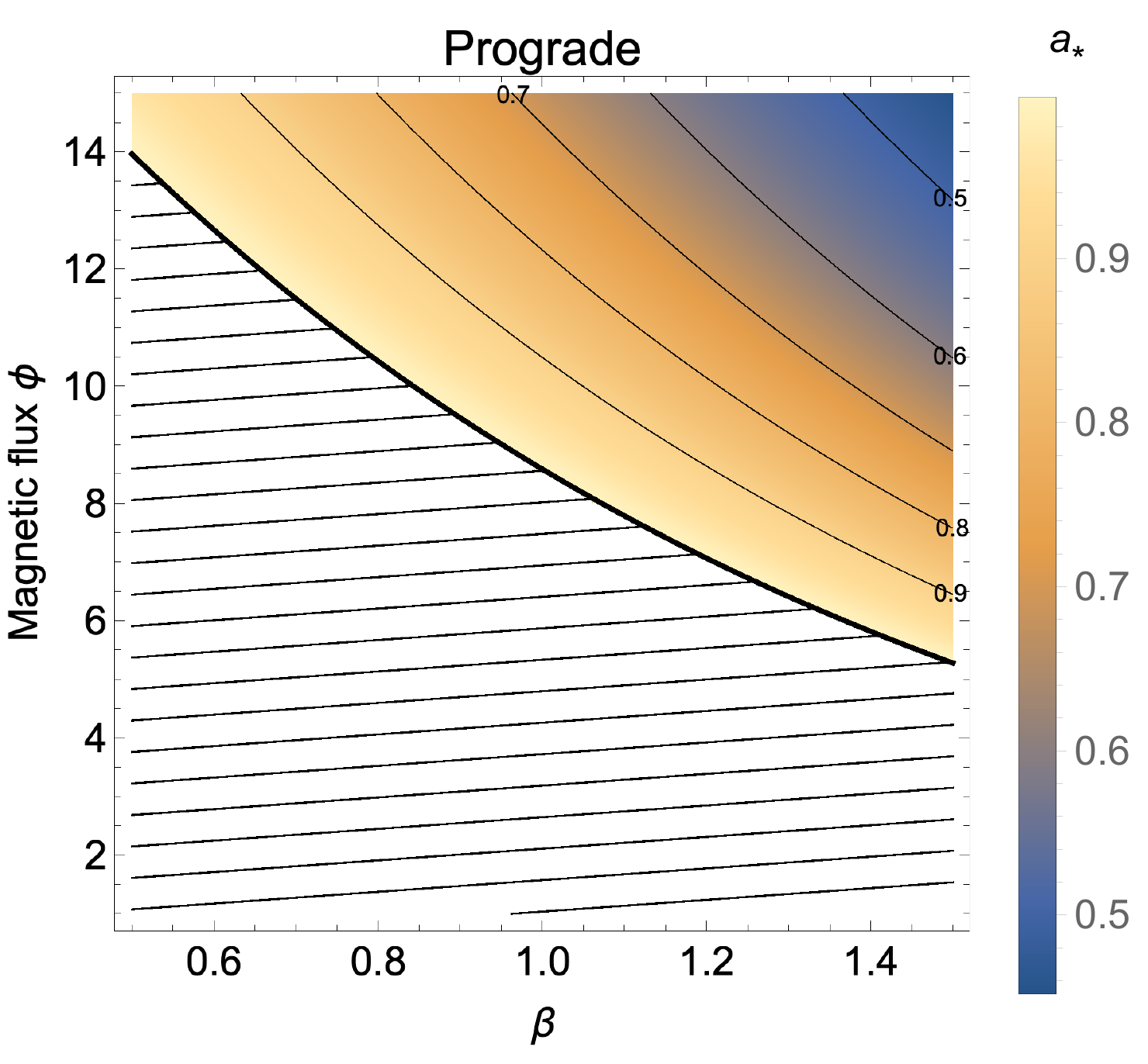} % first figure itself
    \end{minipage}\hfill
    \begin{minipage}{0.45\textwidth}
        \centering
        \includegraphics[width=\textwidth]{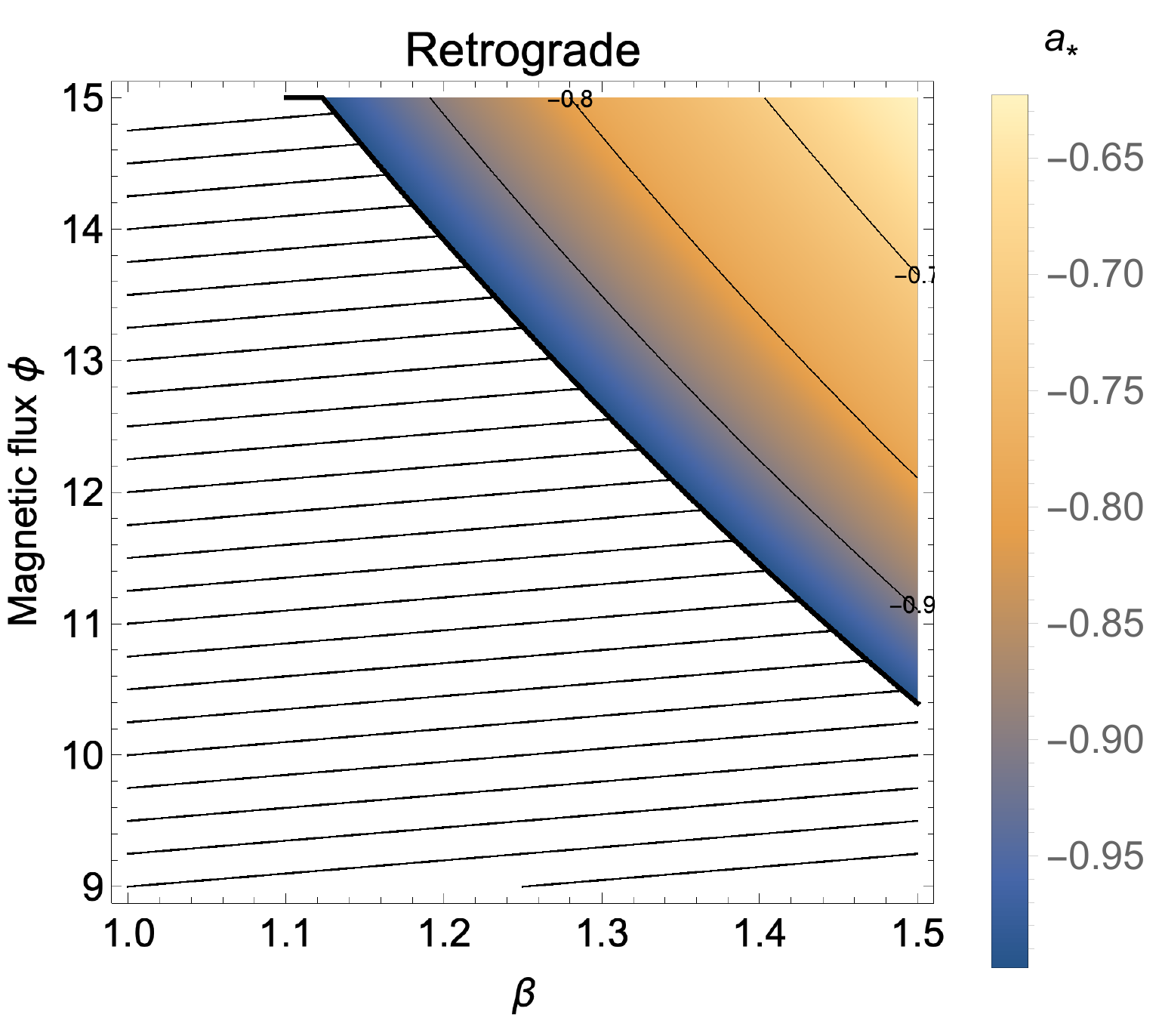} % second figure itself
    \end{minipage}
\caption{Observational parameter space for M87*--spin, magnetic flux and $\beta$--consistent with the average observed jet power. The left and right panels display the prograde and retrograde SMBH cases, respectively. The MAD and SANE states correspond to the top and bottom part of the plots, respectively. The hatched region indicates parameters which are forbidden to the SMBH because they would imply  $|a_*|>0.998$. }
\label{spins-phi}
\end{figure*}

Figure \ref{spins-phi} shows the solutions of the equation \ref{eq} for $a_*$ and $\phi$ that are consistent with the mean values of $\eta_{\rm observed}$ (i.e. the values along the solid line in Fig. \ref{obs-eta}). %, for different possibilities of density profiles parameterized as a range of values of $\beta$. 
As such, Fig. \ref{spins-phi} gives us the observational constraints on M87*'s spin and magnetic flux. To begin with, the hatched area in the plot indicates the region of the parameter space which is forbidden for M87* on the assumption of the Kerr metric, because it would imply $|a_*| > 0.998$ which is the astrophysical limiting value of the spin \citep{Thorne1974}. In other words, in the hatched region the SMBH does not provide enough energy to power the observed jet. We now describe separately the prograde and retrograde cases. The MAD state corresponds to the top part of the plot ($\phi \approx 15$) while the SANE mode with $\phi$ close to one--as considered among the models in EHTC5--corresponds to the bottom region. \\
\textbf{Prograde.} In this case, not all magnetic fluxes are accessible to the SMBH. For instance, only accretion flows with $\phi \gtrsim 5$ are permitted. In the extreme situation that $\beta=0.5$, the RIAF must be in the MAD state. All values of $\beta$ are allowed. \\
\textbf{Retrograde.} Retrograde BHs produce less powerful jets, therefore if the SMBH is retrograde this implies tighter constraints on M87*'s parameters compared to the prograde case. Only RIAFs with  $\phi \gtrsim 10$ and $\beta \gtrsim 1.1$ are possible.  \\

\section{Discussion} \label{sec:disc}

Our results are fundamentally based on the modeling of $\eta_{\rm observed}$--the observationally constrained jet production efficiency of M87*. We compute $\eta_{\rm observed}$ from the jet power measured from the X-ray cavities and BH accretion rate upper limit measured from radio polarization. From our modeling of $\eta_{\rm observed}$, we rule out a considerable region of the BH accretion parameter space for M87*. We have found that most SANE models with $\phi \lesssim 5$--i.e. the family of SANE models considered in EHTC1, EHTC5--are inconsistent with the jet energetics since their magnetic flux gives very small jet efficiencies. Most of the MAD models considered in EHTC5 are consistent with our $\phi$-constraints.
Reassuringly, the MAD models with $a_*=-0.94$ and $\phi=8.04$ rejected by EHTC5 because they fail to produce stable images are also rejected in our work. Further constraints on $a_*$ and $\phi$ can be obtained by using the image scoring in EHTC5 as a prior to the modeling performed here. This deserves further investigation in the future. 

The spin bounds from this work encompass--but are less restrictive--than previous estimates in the literature based on semi-analytic spectral fits \citep{Feng2017}, TeV pair production \citep{Li2009} and jet wobbling \citep{Sobyanin2018}. 

It is interesting to discuss one notable difference between our results and those of EHTC5. The SANE models with $\phi \approx 1$ and $a_*=-0.94$ simulated by EHTC5 are characterized by low jet efficiencies $\eta = 5 \times 10^{-3}$, therefore they do not agree with $\eta_{\rm observed} > 0.08$ of this work. However, they produce jet powers in the range $P \sim 10^{42}-10^{43} \ {\rm erg \ s}^{-1}$ and pass the ``jet power consistency test'' of EHTC5. What is the origin of this apparent contradiction between the simulated large $P$ and low $\eta$ for these models? 
The answer is that $\dot{M}$ in the EHTC5 models is adjusted to produce the observed compact mm flux and is not bound by the upper limit provided by the RM of \cite{Kuo2014}--which we respect in our models. The exact value of $\dot{M}$ in the general relativistic ray-tracing simulations employed in EHTC5 depends on the electron thermodynamics and spans a wide range.

M87*'s mass accretion rate upper limit is obtained using a model for the density and magnetic field strength and geometry in the RIAF in which it is roughly spherical with $\rho(r) \propto r^{-\beta}$ and the magnetic field is well ordered, radial and of equipartition strength (cf. \citealt{Marrone2006,Kuo2014}). There are reasons to believe that the mapping between RM and $\dot{M}$ may not be as straightforward as the prediction by the simple model above \eg{Moscibrodzka2017a}. For instance, since M87's jet has a low viewing angle \eg{Walker2018} it is not impossible that the line of sight does not pass through the accretion flow and the RM measured by Kuo et al. comes from the jet sheath. If this is the case, the RM would be consistent with accretion rates potentially much larger than considered by Kuo et al. \citep{Moscibrodzka2017a}. In the extreme case that $\dot{M}$ is 100 times larger than the Kuo et al. value and assuming our fiducial $P$, then we find $|a_*| > 0.1$, $\phi > 0.5$ in the prograde case and  $|a_*| > 0.15$, $\phi > 1$ in the retrograde case.

What would happen if the current value of $P$ is different than the measurement we have adopted due to e.g. variability? If $P$ is on the lower end of the estimates at $10^{42} \ {\rm erg \ s}^{-1}$ \eg{Prieto2016} then the lower limits on $a_*$ and $\phi$ are somewhat alleviated: $|a_*| > 0.2$, $\phi > 2$ in the prograde case and $|a_*| > 0.3$, $\phi > 4$ in the retrograde one. On the other hand, if the actual $P$ lies on the high end of the estimates at $10^{45} \ {\rm erg \ s}^{-1}$ \eg{Degasperin2012} then the BH is unable to provide enough power to the jet via the Blandford-Znajek process given the $\dot{M}$ measured by \cite{Kuo2014}. The maximum jet power that can be extracted from M87* fed at this $\dot{M}$ is $6 \times 10^{43} \ {\rm erg \ s}^{-1}$, similarly to the most powerful jets in the EHTC5 models. If $P$ is established to be as high as $10^{45} \ {\rm erg \ s}^{-1}$, then either the $\dot{M}$ measurement of Kuo et al. is innacurate or there is a different mechanism (not Blandford-Znajek) powering the jet.

One assumption of our work is that the dependency of $\eta$ on the BH parameters is adequately described by the time-average of global ideal GRMHD simulations of relativistic jet formation. These models have achieved an impressive level of sophistication in the last few years and the results of different GRMHD codes broadly agree with each other \eg{Porth2019}. However, simulations which incorporate kinetic effects are suggesting that jet formation can be more complex than previously thought, with a significant population of  particles with negative energies measured by a far observer that could lead to strong extraction of the BH spin energy via the Penrose process \citep{Parfrey2019}. This could alter expectations for the radiation spectrum from BH accretion. Furthermore, we have only considered the case of a BH spin vector parallel to the angular momentum vector of the accretion flow. The general case of arbitrary relative orientations of these vectors can lead to more complicated jet behaviors \citep{Liska2018}. These issues deserve further investigations. 

Finally, as we already mentioned the BH shadow is weakly affected by the different values of $a_*$ and it is not clear whether further EHT observations will eventually be able to constrain the ring properties with enough sharpness to meaningfully constrain the spin. We are looking forward to advancements in VLBI techniques that will allow better measurements of the ring properties and  
potentially advance our understanding of M87*'s spin (e.g. going to shorter wavelengths, adding more telescopes, going to space-based interferometry; e.g. \citealt{Roelofs2019}) and also to upcoming polarimetric analysis of the EHT observations which will further constrain the magnetic flux and accretion rates.

\section{Summary}	\label{sec:summary}

We have compared measurements of the power carried by particles in the relativistic jet and the black hole mass accretion rate with the predictions of GRMHD models of jet formation, which allows us to derive constraints on the spin and magnetic flux of M87*'s supermassive black hole. The jet power comes from X-ray observations and corresponds to the time-average over $\sim 10^6$ years, while the accretion rate is estimated assuming the Faraday rotation measures come from external Faraday rotation due to the RIAF. Our main conclusions can be summarized as follows:

(i) The black hole in M87* is converting at least $\eta=10\%$ of the accreted rest mass energy to jet power, and up to $200\%$ depending on how shallow the density profile of the accretion flow is.

(ii) We derive a lower limit on M87*'s spin: $|a_*| \geq 0.4$ in the prograde case or $|a_*| \geq 0.5$ in the retrograde case. We are not able to distinguish between the prograde or retrograde scenarios based only on the data we have used.

(iii) We obtained lower limits on the BH magnetic flux, potentially ruling out a variety of models with low values of $\phi$ known as ``SANE''. We find that $\phi \gtrsim 5$ in the prograde case and $\phi \gtrsim 10$ in the retrograde case. Therefore, the magnetically arrested disk state seems to be preferred by M87*.

A possibility that cannot currently be excluded is that the RM could be jet-dominated, given the low inclination angle to the relativistic jet. In this case, the accretion rate could potentially be much larger than we considered. This would be consistent with low spins ($|a_*| \geq 0.1$) or SANE accretion ($\phi \gtrsim 1$). 

We hope that these constraints on the M87*'s black hole spin and magnetic flux will be useful in further extracting physical parameters from M87*'s BH shadow.

\acknowledgments

We would like to thank Sasha Tchekhovskoy for providing us with the electronic data for his GRMHD simulations in a convenient format. We thank Fabio Cafardo, Gustavo Soares and Sasha Tchekhovskoy for useful discussions, and the anonymous referee for a constructive report that led to improvements in the paper.
This work was supported by FAPESP (Funda\c{c}\~ao de Amparo \`a Pesquisa do Estado de S\~ao Paulo) under grant 2017/01461-2. The Black Hole Group at USP acknowledges the generous donation of a GPU from the NVIDIA Corporation. % under the GPU Grant Program.

\vspace{5mm}
\facilities{\textit{Chandra} X-ray Observatory, Submillimeter Array, Event Horizon Telescope}
\software{Jupyter \citep{ipython}, \href{https://github.com/tisimst/mcerp}{\code{mcerp}},   \href{https://github.com/rsnemmen/nmmn}{\code{nmmn}}} % specifies which programs were used

\bibliography{refs}

\end{document}

%% file: definitions.tex
\usepackage[shadow]{todonotes}	% \todo{}: side notes at the margins; \todo[inline]{}: inline notes; \missingfigure{}; \listoftodos
\usepackage{color, soul}	% Highlighting, use \hl{...}
\usepackage{wrapfig}	% wrap text around figures; https://en.wikibooks.org/wiki/LaTeX/Floats,_Figures_and_Captions
\usepackage{verbatim}	% comment out chunks of text
\usepackage[normalem]{ulem} % fancy underlining, strikethrough spanning pages. \sout strikethrough, \xout, \underline

\definecolor{boxback}{HTML}{DAE5F0}
\definecolor{boxframe}{HTML}{0F365B}
{
\end{tcolorbox}   
}

\newcommand{\eg}[1]{(e.g. \citealt{#1})}

\def\code#1{\texttt{#1}}